Stretched Exponential Decay in the Edwards-Wilkinson Model

S. F. Edwards

Cavendish Laboratory

Madingley Rd.

CB3 OHE, Cambridge

U.K.

M. Schwartz

School of Physics and Astronomy

Tel Aviv University

Ramat Aviv, Tel Aviv 69978

Israel

We consider the exactly soluble Edwards-Wilkinson Model in one dimension and demonstrate explicitly, that it is possible to construct a field $\phi(\vec{r})$, that does not depend explicitly on time, such that the corresponding time dependent correlation function, $\chi_q(t) = <\phi_q(t)\phi^*_{-q}(t)>$, is dominated at long times by a stretched exponential decay. The difference between this and the stretched exponential decay present in truly non linear systems is discussed.

We have argued recently,[1-3] that stretched exponential decay is much wider spread in condensed matter physics than formerly expected. In fact, we claim, that such a form of decay is present in many non linear systems such as, the KPZ dynamical systems, ferromagnets at their critical point etc.. These systems have in common the property that a disturbance of wave vector $\vec{q}$, decays at equilibrium with a characteristic decay rate, $\omega_q$, that is proportional to $q^\mu$ with $\mu > 1$.

The purpose of the present article is to demonstrate explicitly that this is also true for linear systems and work out explicitly the form of decay for certain correlation function in the Edwards-Wilkinson[4] system. Because the system is exactly soluble the full form of the decay function can be written in closed form.

The Edwards-Wilkinson model is the simplest model describing the landing of material and its rearrangement on a growing surface. If we denote the height at a point $\vec{x}$ relative to the spatial averaged height by $h(\vec{x})$, the model is described in one dimension by the Langevin equation

$$\frac{\partial h}{\partial t}(x,t) = \nu \nabla^2 h(x,t) + \eta(x,t),  \qquad (1)$$

where the noise, $\eta$, obeys

$$<\eta(x,t)> = 0 \; ; \; <\eta(x,t)\eta(x',t')> = 2D\delta(x-x')\delta(t-t') . \qquad (2)$$

The Langevin equation can be transformed into a Fokker-Planck equation for the probability density, $P\{h_q\}$, to find the Fourier transform $h_q$, of $h(x)$.

$$\frac{\partial P}{\partial t} = \sum \frac{\partial}{\partial h_q}[D\frac{\partial}{\partial h_{-q}} + vq^2 h_q]P. \qquad (3)$$

Using the standard transformation,[5-8] $P = \exp[-\frac{1}{4}\sum \frac{vq^2}{D}h_q h_{-q}]\,\psi$, we obtain for $\psi$ a Schroedinger like equation

$$\frac{\partial \psi}{\partial t} = -H\psi, \qquad (4)$$

where the "Hamiltonian"

$$H = \sum D\{-\frac{\partial^2}{\partial h_q \partial h_{-q}} + \frac{1}{4}(\frac{vq^2}{D})^2 h_q h_{-q} - \frac{1}{2}\frac{vq^2}{D}\} \qquad (5)$$

is Hermitian non-negative definite and its only eigenfunction with eigenvalue zero is $\psi_G \propto \exp[-\frac{1}{4}\sum \frac{vq^2}{D}h_q h_{-q}]$. It proves useful to transform to creation and destruction operators by writing

$$h_q = [\frac{D}{vq^2}]^{\frac{1}{2}}(\alpha_q + \alpha_{-q}^+) \qquad (6)$$

and

$$\frac{\partial}{\partial h_q} = -\frac{1}{2}[\frac{D}{vq^2}]^{-\frac{1}{2}}(\alpha_q^+ - \alpha_{-q})$$

(7)

where the $\alpha's$ obey the usual Bose commutation relations

$$[\alpha_q, \alpha_p] = [\alpha_q^+, \alpha_p^+] = 0 \ ; \ [\alpha_q, \alpha_p^+] = \delta_{qp} \ .$$  (8)

In terms of these operators,

$$H = \sum vq^2 \alpha_q^+ \alpha_q \ ,$$  (9)

namely a Hamiltonian of free excitations

Consider next a general composite field $\phi(x)$. Its time dependent structure factor is

$$\Gamma_q(t) = <\phi_{-q}^*(0)\phi_{-q}(t)> \ ,$$  (10)

where the meaning of the average above is that $\phi_q$ is measured at steady state, the system is then allowed to evolve freely and then after a time $t$, $\phi_{-q}$ is measured. It is a standard procedure to show that

$$\Gamma_q(t) = <0|\phi_q \ e^{-Ht} \phi_{-q}^*|0> \ ,$$  (11)

here the vacuum $|0>$ is the normalized state with eigenvalue zero of $H$. Introducing eigenstates of $H$ with momentum $q$, we obtain the general result,

$$\Gamma_q(t) = \sum_\beta |<0|\phi_q|q,\beta>|^2 \ e^{-\lambda_{q\beta} t}, \tag{12}$$

where $|q,\beta>$ is a normalized eigenfunction of $H$ that is also an eigenfunction of the momentum operator with eigenvalue $q$ and $\lambda_{q\beta}$ is the corresponding eigenvalue of $H$.

It is clear from eq. (12), that a necessary condition for a slower than exponential decay of $\Gamma_q(t)$ is that for any $q$ the spectrum $\lambda_{q\beta}$ is gapless.

In our case

$$|q,\beta> \equiv |\{n_\ell\}> \quad \text{(such that } \sum \ell n_\ell = q\text{)}, \tag{13}$$

where $n_\ell$ is the occupation number of the state $\ell$. The corresponding eigenvalue is

$$\lambda_{q,\beta} = \sum v\ell^2 n_\ell \quad \text{(such that } \sum \ell n_\ell = q\text{)}. \tag{14}$$

Consider the state with $n_{q/n} = n$ and $n_\ell = 0$ for any other value of $\ell$. The corresponding eigenvalue is

$$\lambda_q^{(n)} = v\frac{q^2}{n}. \tag{15}$$

Since $n$ can take any integer value it is obvious that the spectrum $\lambda_{q,\beta}$ is indeed gapless and the necessary condition for slow decay is fulfilled.

Now consider for example the case where $\phi(x) = h(x)$. It is obvious that the decay is not slow but actually $\Gamma_q(t) = \frac{D}{vq^2} e^{-vq^2 t}$. The reason is that the only matrix element of the form $<0|h_q|q,\beta>$ that is not zero is the state with $n_q = 1$ and all other $n_\ell$'s being zero. If we take $\phi(x) = h^n(x)$ (recall that $h$ does not have a zero Fourier transform) we arrive again at the conclusion that the decay of $\Gamma_q(t)$ is exponential, because the matrix elements do not couple with states that have a vanishingly small "energy". In fact, if $\phi(x) = h^n(x)$ the decay will be not slower than $e^{-v\frac{q^2}{n}t}$. The conclusion is that in order to obtain a slow decay we must consider such a composite field that will couple the vacuum to states with arbitrarily small "energies".

We consider here the example $\phi(x) = \exp[i\alpha h(x)]$, where obviously $\alpha$ is a constant that has the dimensions of $h^{-1}$. It is straightforward to show that for such a composite field

$$<\phi_q^*(0)\phi_{-q}(t)> = \int dx \; \exp[-\frac{1}{2}\alpha^2 w(x,t) - iqx], \qquad (16)$$

where

$$w(x,t) = <[h(x,t) - h(0,0)]^2> = D[\frac{t}{v}]^{\frac{1}{2}} f(\frac{x}{\sqrt{vt}}) \;. \qquad (17)$$

The time dependent correlation function can be written now in the form

$$\Gamma_q(t) = \sqrt{vt} \int dy \; \exp[-\alpha^2 D(\frac{t}{v})^{\frac{1}{2}} f(y)] \; e^{-iq\sqrt{vt}\, y} \;. \tag{18}$$

We are interested in the behaviour of $\Gamma_q(t)$ for $qv/\alpha^2 D \ll 1$ and $vq^2 t \gg 1$. It is easy to show that under these conditions only small values of $y$ are relevant to the integration. The conclusion is

$$\Gamma_q(t) = A\sqrt{\frac{2v^{3/4}}{\alpha^2 D}} \; t^{\frac{1}{4}} \; \exp[-\frac{2v^{3/2} q^2 t^{\frac{1}{2}}}{\alpha^2 D \sqrt{\pi}}], \tag{19}$$

where $A = \int_{-\infty}^{\infty} \frac{dy}{y^2}[1 - e^{-y^2}] \;.$

We see here that the fact that the spectrum is gapless for any $q$ manifests itself in a decay that is dominated by a stretched exponential in time, Note, however, that the structure is not of the form $\tilde{\Gamma}_q f(\omega_q t)$ as is expected in the non linear problem. The reason is that in the definition of the field $\phi(x)$ we introduced a dimensional constant $\alpha$ and as a result a length, $x_0$, into the problem that does not appears in the physics. The explicit dependence of the correlation function on the new length $x_o = v/\alpha^2 D$ is expressed by the following form of $\Gamma_q$,

$$\Gamma_q(t) = \frac{A2^{1/2}}{q}[x_0 q]^{1/2}(\omega_q t)^{1/4}\exp[-\frac{2}{\sqrt{\pi}}(x_0 q)(\omega_q t)^{1/2}] \ . \tag{20}$$

This long time decay is very different from the decay of the function $<h_q(t)h_{-q}(0)>$ in non linear systems of the KPZ nature or of dynamical $\phi^4$ theory at the transition , that is also found to be dominated by stretched exponential decay.

Here we have no restriction on the $q$ dependence of the exponential. In the non-linear systems, we studied[1,2,] we found that the non linearity forces that the exponential describing the decay of say $<h_q(t)h_q(0)>$ or analogues of it, to be linear in $|q|$ and since the dependence of the exponential on time came only through the combination $\omega_q t$, it follows that the stretched exponential factor must be of the form $\exp[-c|q|t^{1/\mu}]$[9].